\documentstyle[aps,multicol,prl,epsfig]{revtex}

\newcommand{\beq}{\begin{equation}}
\newcommand{\eeq}{\end{equation}}

\begin{document}

\begin{title}
{\bf Dark-in-Bright Solitons in Bose-Einstein Condensates with 
Attractive Interactions}
\end{title}

\author{P.G. Kevrekidis$^{1}$, D.J. Frantzeskakis$^{2}$, Boris A. Malomed$^3$, 
A.R. Bishop$^4$, and I.G. Kevrekidis$^{5}$}
\address{$^{1}$ Department of Mathematics and Statistics,University of Massachusetts,
Amherst MA 01003-4515, USA \\
$^{2}$ Department of Physics, University of Athens,
Panepistimiopolis,Zografos, Athens 15784, Greece \\
$^{3}$ Department of Interdisciplinary Studies, Faculty of Engineering, 
Tel Aviv University, Tel Aviv 69978, Israel \\
$^{4}$ Center for Nonlinear Studies and Theoretical Division, Los
Alamos National Laboratory, Los Alamos, NM 87545 USA \\
$^5$ Department of Chemical Engineering and Program in Applied and
Computational Mathematics, 6 Olden Street, Princeton University,
Princeton, NJ 08544, USA}
\date{\today}
\maketitle

\begin{abstract}
We demonstrate a possibility to generate localized states in effectively
one-dimensional Bose-Einstein condensates with a negative scattering length
in the form of a dark soliton in the presence of an optical lattice (OL)
and/or a parabolic magnetic trap. We connect such structures with twisted
localized modes (TLMs) that were previously found in the discrete nonlinear
Schr{\"o}dinger equation. Families of these structures are found as
functions of the OL strength, tightness of the magnetic trap, and chemical
potential, and their stability regions are identified. Stable bound states
of two TLMs are also found. In the case when the TLMs are unstable, their
evolution is investigated by means of direct simulations, demonstrating
that they transform into large-amplitude fundamental solitons. An analytical
approach is also developed, showing that two or several fundamental
solitons, with the phase shift $\pi$ between adjacent ones, may form stable
bound states, with parameters quite close to those of the TLMs revealed by
simulations. TLM structures are found numerically and explained analytically
also in the case when the OL is absent, the condensate being confined only
by the magnetic trap.
\end{abstract}

\vspace{2mm}

\section{Introduction}

The current experimental and theoretical studies of Bose-Einstein
condensates (BECs) \cite{review} have attracted a great deal of interest to
nonlinear patterns which can exist in them, including dark \cite{dark} and
bright \cite{bright} solitons. Two-dimensional (2D) excitations in the form
of vortices were also realized experimentally \cite{vortex}. Other nonlinear
excitations, such as Faraday waves \cite{stal}, ring dark solitons and
vortex necklaces \cite{theo} were predicted to occur in BECs.

The behavior of the condensate crucially depends on the sign of the atomic
interactions: dark (bright) solitons can be created in BECs with repulsive
(attractive) interactions, resulting from the positive (negative) scattering
length. Recent experiments have demonstrated that ``tuning'' of the
effective scattering length, including a possibility to change its sign, can
be achieved by means of a Feshbach resonance \cite{inouye}.

In this work we demonstrate that another type of localized nonlinear
excitations, which may be regarded as dark solitons embedded in bright ones,
can be created in attractive BECs. They may also be considered as bound
states of two bright solitons with a phase shift $\pi $ between them. In
some respects, we follow the path of very recent results of Ref. 
\cite{louis}. These, in turn, were based on earlier works on the 
so-called twisted
localized modes (TLMs) in discrete systems \cite{DKL}; that is why we apply
the same term, TLM, to these solitons of the combined dark-bright type. TLMs
in a discrete system are bright solitons in which the field vanishes,
changing its sign, at the central point. An alternative way to view
a TLM is as a concatenation of an ``up-pulse'' (i.e., ${\rm sech}(x-x_1)$)
with a ``down pulse'' (i.e., $-{\rm sech}(x-x_2)$), where $x_1$ and $x_2$
are appropriately separated, to form an up-down, two-pulse configuration. 
Notice, however, that our work is essentially
different from Ref. \cite{louis}, since we focus on the case of BECs with 
{\it negative} scattering length (such as $^{7}$Li \cite{Li} and $^{85}$Rb 
\cite{Ru}), and demonstrate robustness and stability of TLMs in the latter
context. These structures, if regarded as effectively dark solitons in the
attractive BEC, are the inverse of recently predicted \cite{vvkbbms} bright
solitons in repulsive BECs with an optical lattice. It is expected that a
characteristic longitudinal length of the TLM soliton in physical units will
be on the order of the wavelength of light which induces the OL, i.e., $\sim
1$ $\mu $m, and the number of atoms in the soliton may typically be $\sim
10^{3}$ -- $10^{4}$.

We consider a quasi-1D (cigar-shaped) BEC with negative scattering length,
the corresponding normalized Gross-Pitaevskii (GP) equation being \cite
{review,rupr,GPE1d} 
\begin{equation}
iu_{t}=-u_{xx}-|u|^{2}u+\left[ \epsilon x^{2}+V_{0}\cos ^{2}\left( 2\pi
x/\lambda \right) \right] u,  \label{tlmeq1}
\end{equation}
where $u(x,t)$ is the macroscopic wave function, $\epsilon $ measures the
strength of the parabolic magnetic trap, and the last term accounts for the
OL potential created by interference of two counterpropagating coherent
light beams \cite{catal,greiner,tromb,konot,tromb2,catal2}. As we consider
only (effectively) 1D configurations, the attractive interactions between
the atoms (characterized by the negative scattering length) can not produce
collapse in the framework of the present model, hence various solitons have
a chance to be stable.

It is obvious that scaling invariance of Eq. (\ref{tlmeq1}) makes it
possible to fix the optical wavelength, setting $k\equiv 2\pi /\lambda =1$.
This essentially means that all the remaining length scales of the
problem are measured by comparison to the period of the optical lattice
which has been fixed to $\pi$.
We follow this convention below. The number of atoms in the condensate,
given by 
\begin{equation}
N=\int_{-\infty }^{+\infty }|u(x)|^{2}dx,  \label{N}
\end{equation}
is a dynamical invariant of Eq. (\ref{tlmeq1}). 

In the next section we report detailed numerical results for TLM solutions
in attractive BECs, as modeled by Eq. (\ref{tlmeq1}). In section III, we
will support some of the numerical findings by analytical considerations.
Section IV concludes the paper.

\section{Numerical results}

\subsection{Stationary twisted-localized-mode solutions and their linear
stability}

We seek stationary solutions to Eq. (\ref{tlmeq1}) in the form, 
\[
u(x,t)=\exp (-i\mu t)v(x),
\]
where $\mu $ is the chemical potential, and a real function $v(x)$ obeys the
equation 
\begin{equation}
\mu v=-v_{xx}-v^{3}+\left[ \epsilon x^{2}+V_{0}\cos ^{2}(2\pi x/\lambda )
\right] v\text{,}  \label{v}
\end{equation}
supplemented with free boundary conditions.  We have checked that the
boundary conditions do not significantly affect the results. The
corresponding boundary-value problem was solved by means of a
finite-difference discretization.
A second order difference scheme with spacing $\Delta x=0.2$ was
used to approximate $v_{xx}$. The resulting set of nonlinear algebraic
equations was solved by means of a Newton iteration.
More details on finite-difference schemes and Newton type methods
can be found in \cite{findif}.
The solutions of interest were obtained by choosing, as the
initial condition for the Newton iterations, the following functional form: 
\begin{equation}
v(x)=\sum_{j=1}^{n}(-1)^{j}\eta {\rm sech}\left( \frac{\eta }{\sqrt{2}}
(x-\xi _{j})\right) ,\eta \equiv \sqrt{-\mu }.  \label{ansatz}
\end{equation}
The ansatz (\ref{ansatz}) approximates a superposition of 
nonlinear-Schr{\"{o}}dinger (NLS) solitons. The presence of 
the OL suggests choosing positions of
the solitons' centers, $\xi _{j}$, as $\xi _{j}=j\pi /2$, with some integer 
$j$. The sign factor $(-1)^{j}$ implies the phase difference $\pi $ between
adjacent solitons. This choice of the phase pattern is suggested by the
known result that, in the discrete NLS equation, only such a configuration
may be stable, see Ref. \cite{kkm} and references therein. Similar results
for a BEC model with elliptic-function potentials were obtained in Ref. \cite
{jared}. The number $n$ of solitons in the initial ansatz (\ref{ansatz}) was
taken as $n=2$, in order to find the TLM soliton proper, or $n=3$, with the
objective to find a bound state of two TLMs, see below. However, it will
become clear below that 
concatenating more of such ``building block'' structures (such as the 
ones with $n=2$ or $n=3$) one can, in principle, construct multi-pulse
configurations with a larger number ($n$) of elementary pulses.

Once TLM states were found as solutions to Eq. (\ref{v}) [starting with the
initial approximation (\ref{ansatz})], their stability was subsequently
analyzed by examining the corresponding linearized problem for small
perturbations. To this end, the perturbed solution was taken as 
\begin{equation}
u(x,t)=e^{-i\mu t}\left\{ v(x)+\delta \left[ a(x)e^{-i\omega
t}+b(x)e^{i\omega ^{\star }t}\right] \right\} ,  \label{ab}
\end{equation}
where $\delta $ is an infinitesimal amplitude of the perturbation, $a(x)$
and $b(x)$ are eigenfunctions of a perturbation mode, and $\omega $ is the
corresponding eigenfrequency; $\ast $ standing for the complex conjugation.
Using the ansatz of Eq. (\ref{ab}) to O$(\delta)$, one obtains
the linearization problem as an eigenvalue problem where $\omega$ is the
corresponding eigenfrequency, while $(a,b^{\star})^T$ (where the $T$ denotes
transpose), is the corresponding eigenfunction. Using the same 
finite difference scheme as the one used for computing  $v(x)$, we convert
this to a matrix eigenvalue problem and find {\it all} the eigenvalues
of the corresponding matrix. This is done by standard matrix eigenvalue
solvers built into Matlab \cite{matlab}. $\omega$ can, in general, be complex,
i.e., $\omega=\omega_r + i \omega_i$, where the subscripts denote
the real and imaginary part of the eigenfrequency.
If eigenfrequencies with non-zero imaginary parts exist, they lead to
exponential instabilities, while their absence implies linear stability of
the solution under consideration. This is monitored by the spectral plane
pictures of the imaginary ($\omega_i$) versus the real ($\omega_r$) part
of the eigenfrequency (see e.g., the bottom right panels of Figs. 
\ref{tlmfig1}-\ref{tlmfig3} below). The presence of even a single
$\omega$ with $\omega_i \neq 0$ implies instability.
In all the cases when stable stationary
solitons were identified according to this criterion (see below), the
stability was also checked and confirmed by direct simulations of the full
equation (\ref{tlmeq1}). For more details on the setup and solution of
the eigenvalue problem, the interested reader is referred to \cite{pgk,JohAub}.

To systematically trace the evolution of numerically exact TLM solutions of
Eq. (\ref{tlmeq1}), we performed one-parameter continuations in the
three-dimensional parameter space of Eq. (\ref{v}), $\left( \mu ,\epsilon
,V_{0}\right)$. Recall that we have set $\lambda \equiv 2\pi $, hence 
$\lambda $ is not a free parameter. We start by varying the OL-potential's
strength $V_{0}$ for fixed $\epsilon =0$ and $\mu =-1.5$. We note that
negative values of the chemical potential are typical of cases when the
corresponding wave-function pattern is localized, although $\mu $ can
sometimes be positive (see below), which is explained by the presence of the
OL potential in Eq. (\ref{v}).

Numerical results corresponding to this case (varying $V_{0}$) are presented
in Fig. \ref{tlmfig1}. In the top subplot of the upper part of the figure,
the number of atoms in the soliton, defined as per Eq. (\ref{N}), is shown
versus $V_{0}$, and the largest instability growth rate, found from the
associated linearized problem, is shown in the bottom panel of the upper
part of Fig. \ref{tlmfig1} (zero instability growth rate implies that the
soliton is stable). Three characteristic examples of the TLM solution are
displayed, together with their linear-stability eigenfrequencies, in the
lower part of Fig. \ref{tlmfig1}.

This solution branch exists only for $V_{0}>0.075$. The presence of this
cutoff point is expected, as the optical lattice is crucial for the
existence of TLM solutions, which are not found in the usual NLS equation
(at least, in the absence of the parabolic-trap potential). Nevertheless,
below we will present a case in which a TLM/dark soliton solution is
possible even in the absence of OL, being supported solely by the magnetic
trap. In the interval $0.275<V_{0}<0.7$, this soliton family is subject to
oscillatory instability which is accounted for by the so-called Hamiltonian
Hopf bifurcation \cite{VdM}. In the latter, the collision of two pairs of
eigenvalues of opposite Krein signature \cite{VdM} results in the generation
of a quartet of genuinely complex eigenvalues. The strongest instability,
with ${\rm Im\,} \omega \approx 0.107$ occurs at $V_{0}\approx 0.5$.

\newpage

\begin{figure}[h]
\vspace{10mm}
\epsfxsize=9cm 
\center{\epsffile{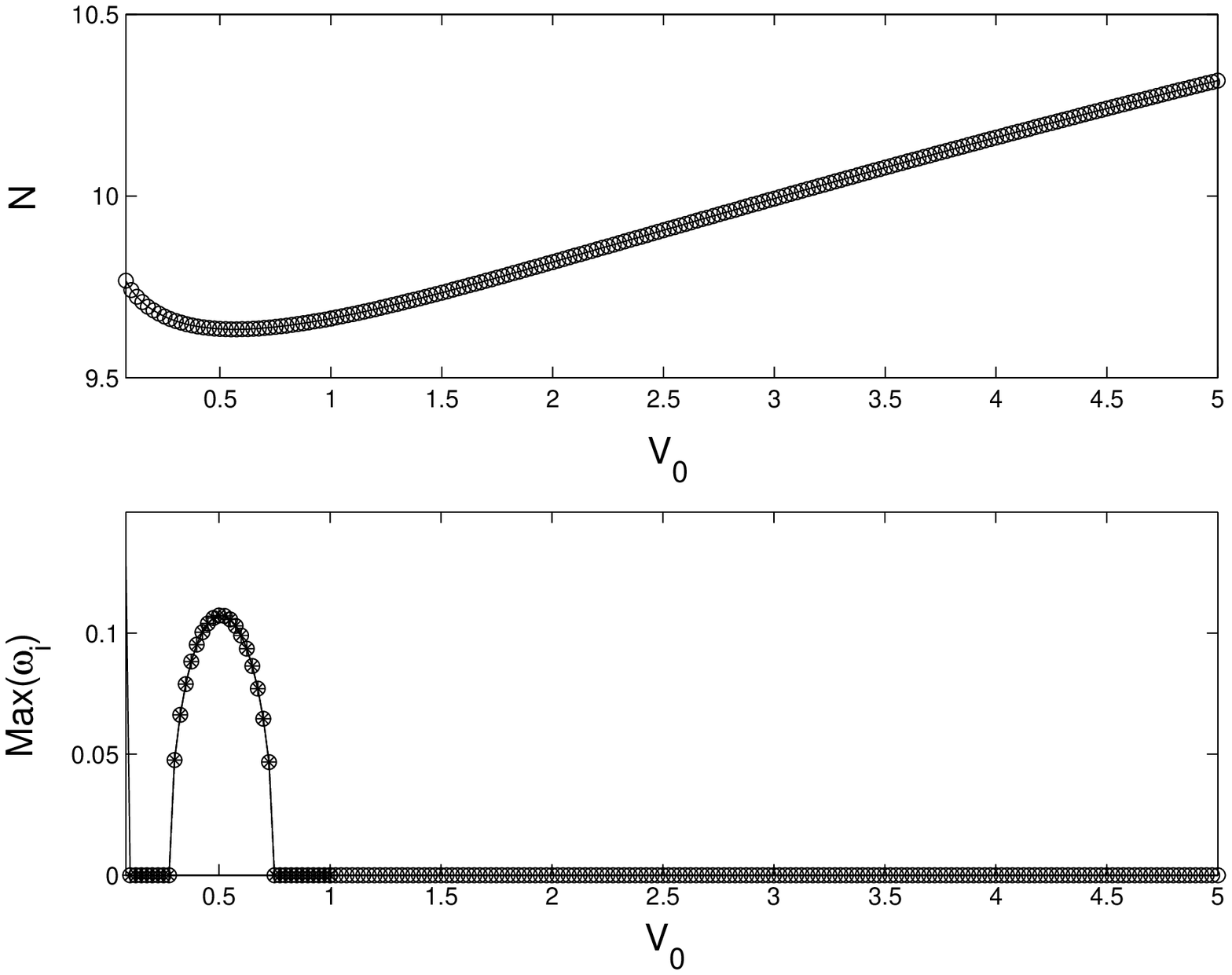}}
\epsfxsize=9cm 
\center{\epsffile{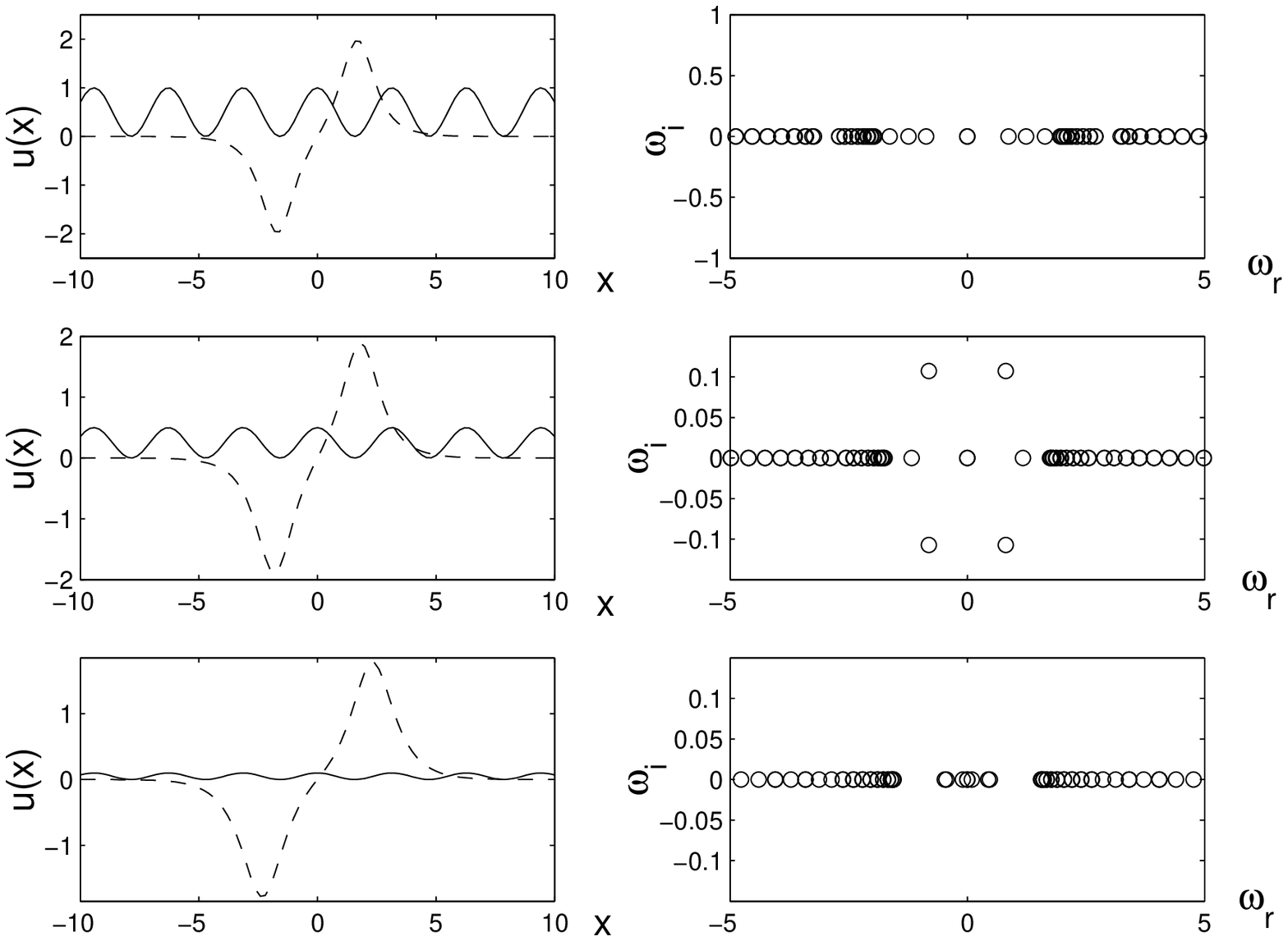}}
\caption{
Upper part: the top panel shows the number of particles vs. the
OL-potential's strength. The bottom panel shows the largest imaginary part
of unstable eigenfrequencies (i.e., the instability growth rate) as a
function of $V_{0}$. Lower part: examples of the TLM solutions for $V_{0}=1$
(top panel), $V_{0}=0.5$ (middle panel) and $V_{0}=0.1$ (bottom panel) are
shown by dashed curves in the left parts of the panels, while the solid line
represents the OL potential. The right panel shows the spectral plane, 
$\omega _{r},\omega _{i}$), of the real and imaginary parts of the
corresponding eigenfrequencies. }
\label{tlmfig1} 
\end{figure}

Similarly, we examined the evolution of the solutions, setting $V_{0}=1$ and 
$\mu =-1.5$ in Eq. (\ref{v}) and performing continuation in $\epsilon $,
starting from $\epsilon =0$. In this case too, oscillatory instability was
found in a finite interval, $0.05<\epsilon <1.61$. The maximum instability
growth rate (which is very large in this case) is $0.48$, and occurs at 
$\epsilon \approx 0.6$. The results for this case are shown in Fig. \ref
{tlmfig2}.

\begin{figure}[h]
\vspace{10mm}
\epsfxsize=9cm 
\center{\epsffile{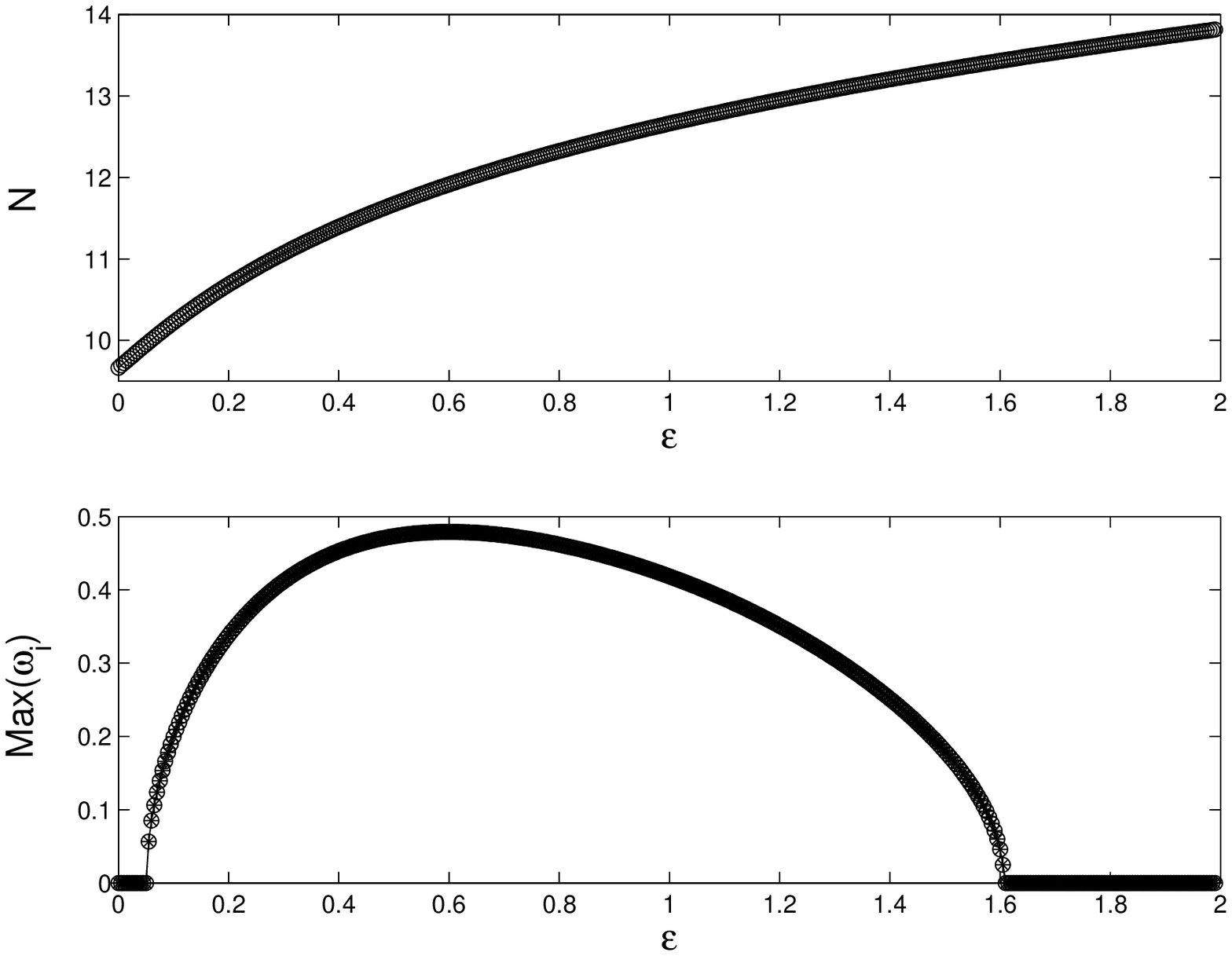}}
\epsfxsize=9cm 
\center{\epsffile{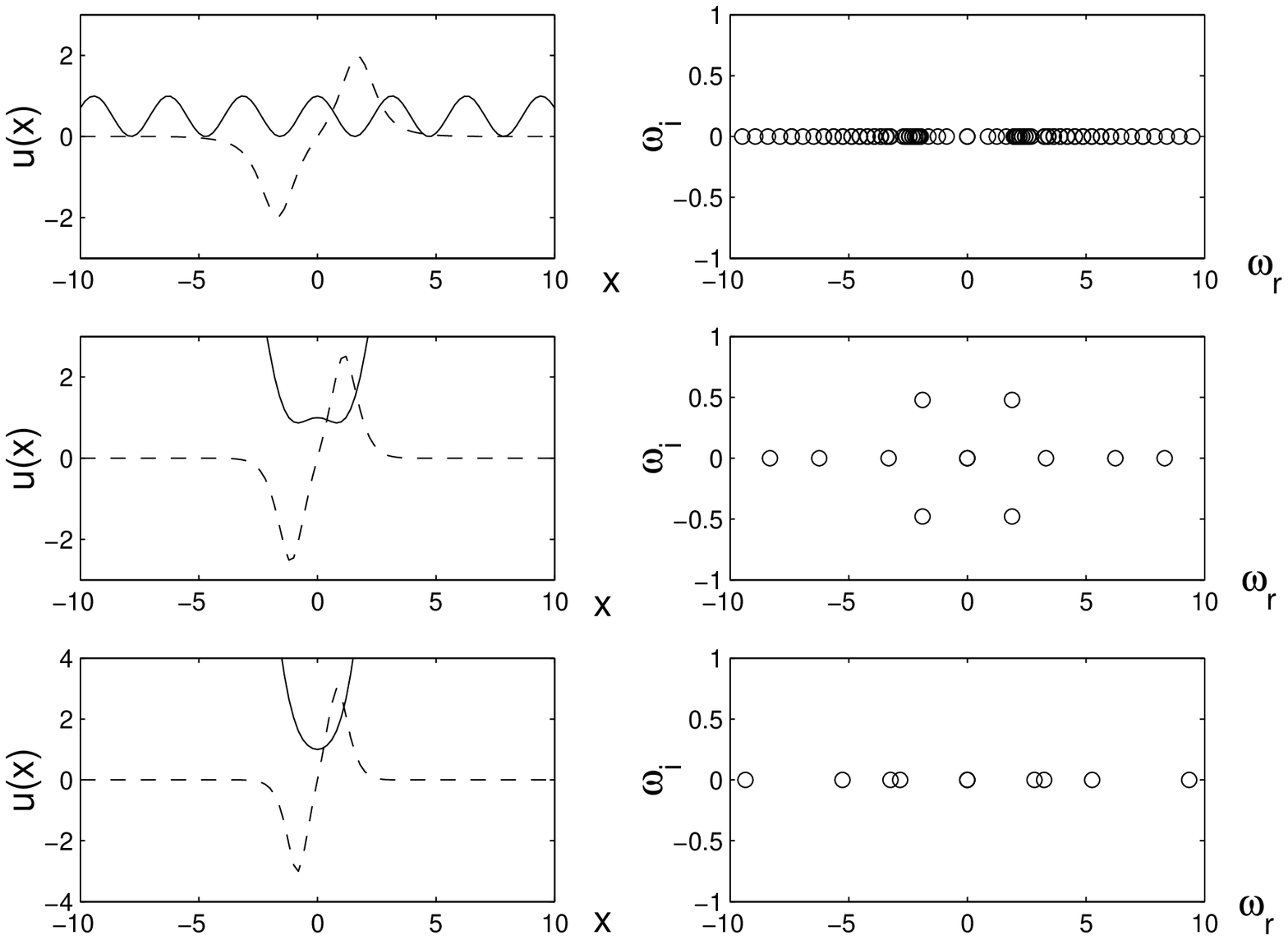}}
\caption{
The same as Fig. 1, but versus the variation of the parabolic-trap strength 
$\epsilon$. In the lower part, three examples of the solutions and their
stability are given for $\epsilon=0$, $\epsilon=0.6$ and $\epsilon=1.75 $.}
\label{tlmfig2} 
\end{figure}

In these cases, the oscillatory instability arises quite naturally. Indeed,
similar instabilities had been observed in the discrete counterpart of the
system and, in fact, in the next section we will give a theoretical
justification for their occurrence. Another interesting numerical finding is
that the TLM solutions appear to be much more robust than one would
initially suspect. In particular, there seem to exist TLM solutions even
when the OL has been almost completely smeared out by the harmonic trap (see
e.g., the last panel of Fig. \ref{tlmfig2}). We will return to this point in
the following section.

We also performed the continuation of the TLM solutions with respect to the
chemical potential $\mu $, which has yielded results displayed in Fig. \ref
{tlmfig3}. In this case, $\epsilon =0$ and $V_{0}=1$ were chosen. The
solutions exist in the region $\mu <\mu _{\max }\approx 0.125$. The fact
that the cutoff value $\mu _{\max }$ is positive is explained by a
contribution of the mean value of the OL potential $V_{0}\cos ^{2}x$ in Eq. 
(\ref{v}). In this case, oscillatory instability is found in two finite
intervals, $-0.375<\mu <-0.1$, and $-0.075<\mu <0$.

\begin{figure}[h]
\vspace{10mm}
\epsfxsize=9cm 
\center{\epsffile{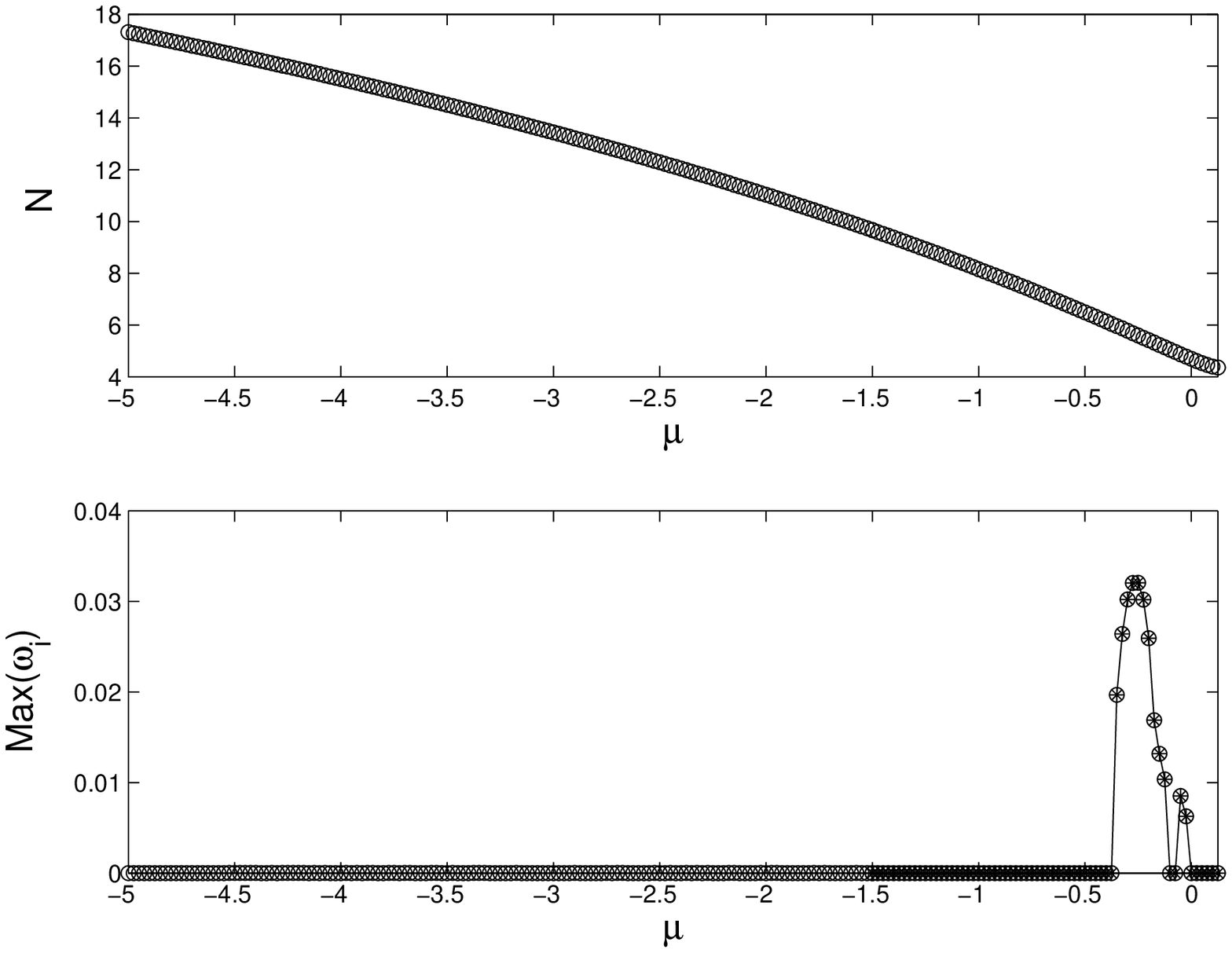}}
\epsfxsize=9cm 
\center{\epsffile{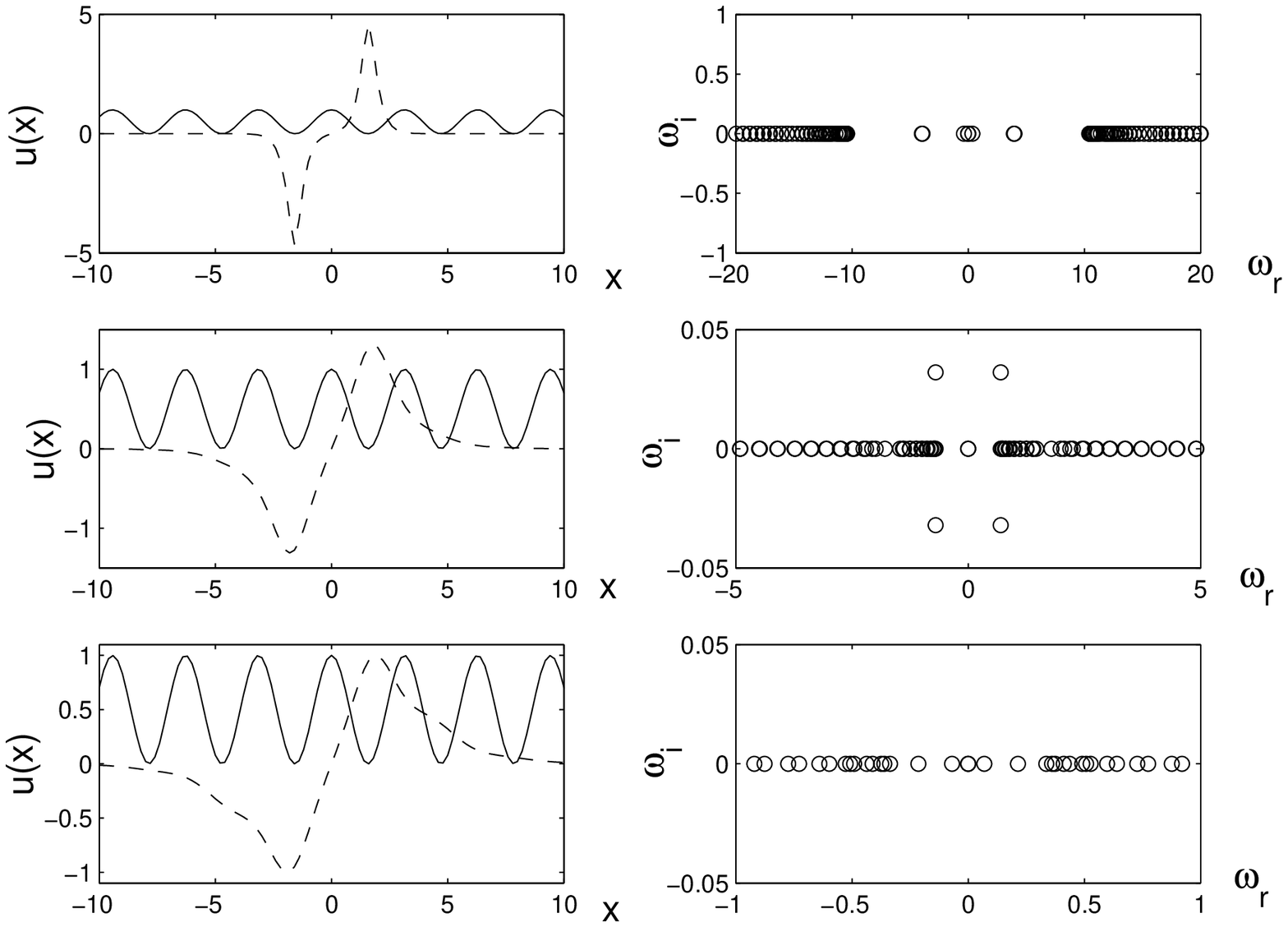}}
\caption{
The same as in Fig. 1 for the variation with respect to the chemical
potential $\mu$. In the lower part, three examples of the TLM solutions
correspond to $\mu=-10$ (top), $\mu=-0.25$ (middle) and $\mu=0.125$
(bottom).}
\label{tlmfig3} 
\end{figure}

\subsection{Bound states of two TLM solitons}

\begin{figure}[h]
\vspace{10mm}
\epsfxsize=9cm 
\center{\epsffile{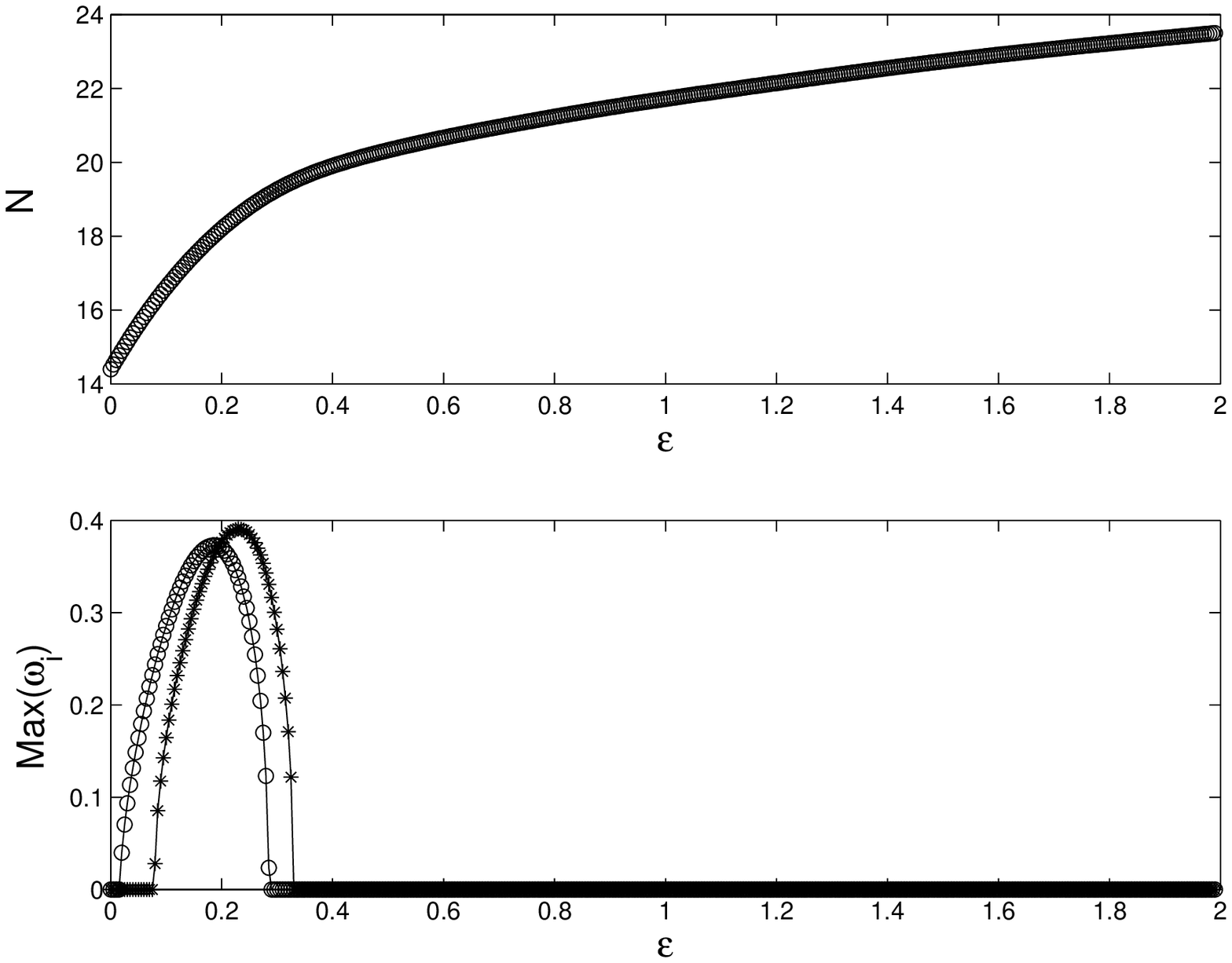}}
\epsfxsize=9cm 
\center{\epsffile{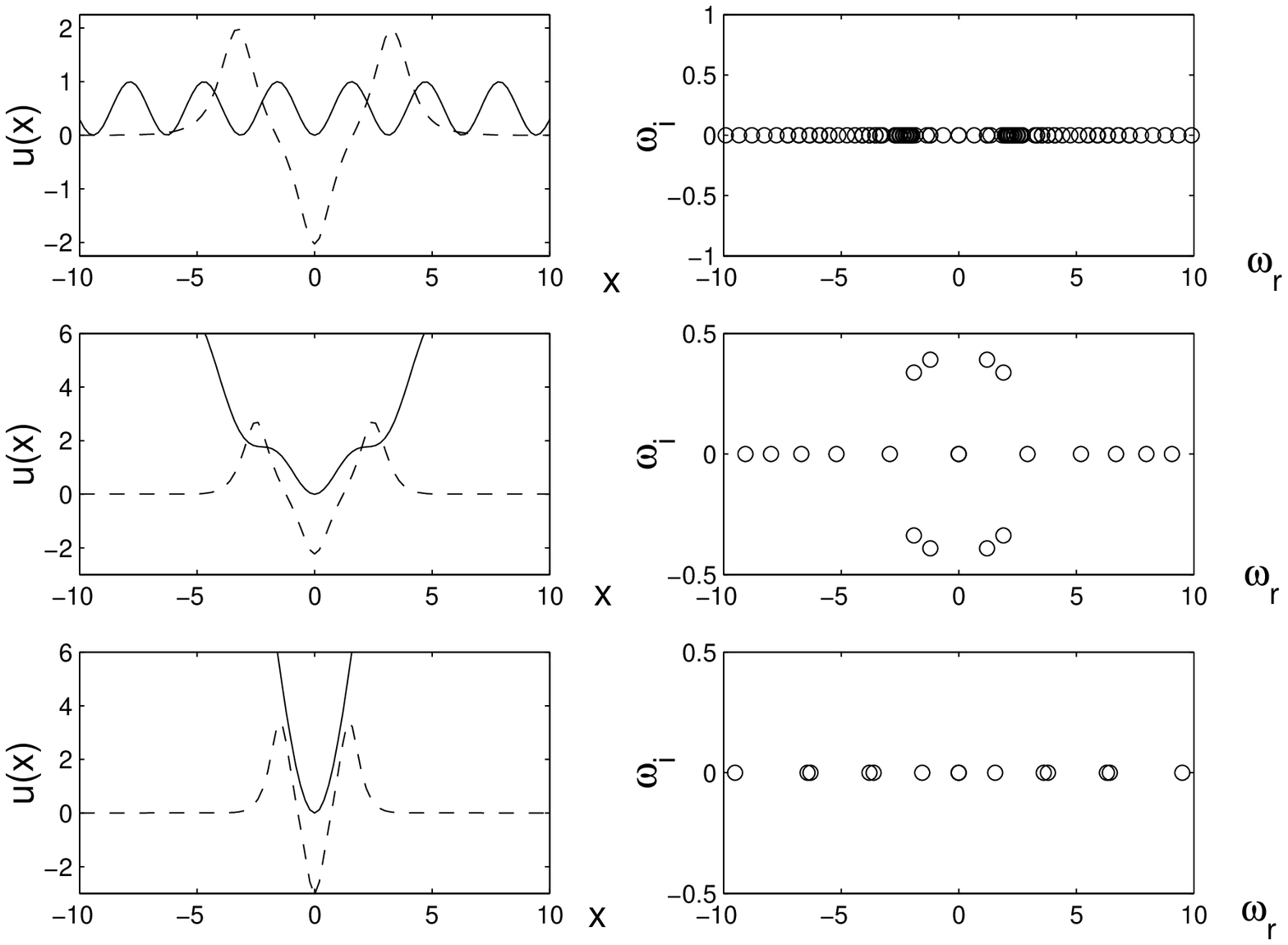}}
\caption{
The same as in Fig. 2, but for the bound state of two TLM solitons. In the
lower part, three examples are shown, for $\epsilon =0$ (top), $\epsilon
=0.23$ (middle) and $\epsilon =2$ (bottom panel).
Notice that in the second panel of the figure two sets of unstable 
eigenfrequencies are shown (one by stars and one by circles). This 
indicates the presence of {\it two unstable eigenvalue quartets}, as 
can also be verified by the spectral plane figure corresponding to
the case of $\epsilon=0.23$; see also the discussion in the text.}
\label{tlmfig4} 
\end{figure}

Complexes which may be considered as a bound state of two TLMs, or simply as
a concatenation of three fundamental solitons with  phase difference $\pi 
$ between adjacent ones, were also studied. In this case, for symmetry
purposes, $\cos x$ in Eqs. (\ref{tlmeq1}) and (\ref{v}) was replaced by 
$\sin x$ (recall we have set $\lambda =2\pi $), and three pulses in the
initial ansatz (\ref{ansatz}) were set at $\xi _{j}=-\pi, 0, \pi $. We
performed the continuation in $\epsilon $, which has demonstrated that the
bound state persists to very large values of $\epsilon $, see Fig. \ref
{tlmfig4}. The branch becomes unstable, through a quartet of complex
eigenfrequencies, at $\epsilon >0.015$. A second instability, accounted for
by another eigenvalue quartet, sets in at $\epsilon >0.075$. This second
oscillatory instability can be easily explained: as we argue below, for each
pair of anti-phase pulses there exists a pair of eigenvalues with negative 
{\it Krein signature}, which is known to give rise to a Hamiltonian Hopf
bifurcation \cite{VdM}. The strongest instability is found at $\epsilon
=0.23 $, when the two instability growth rates are $0.39$ and $0.35$. The
first quartet returns to stability at $\epsilon >0.29$, while the second
quartet is stabilized for $\epsilon >0.33$. The\ bound states are {\em
completely stable} for $\epsilon >0.33$.

\subsection{Evolution of unstable TLM solitons}

\begin{figure}[h]
\epsfxsize=9cm 
\center{\epsffile{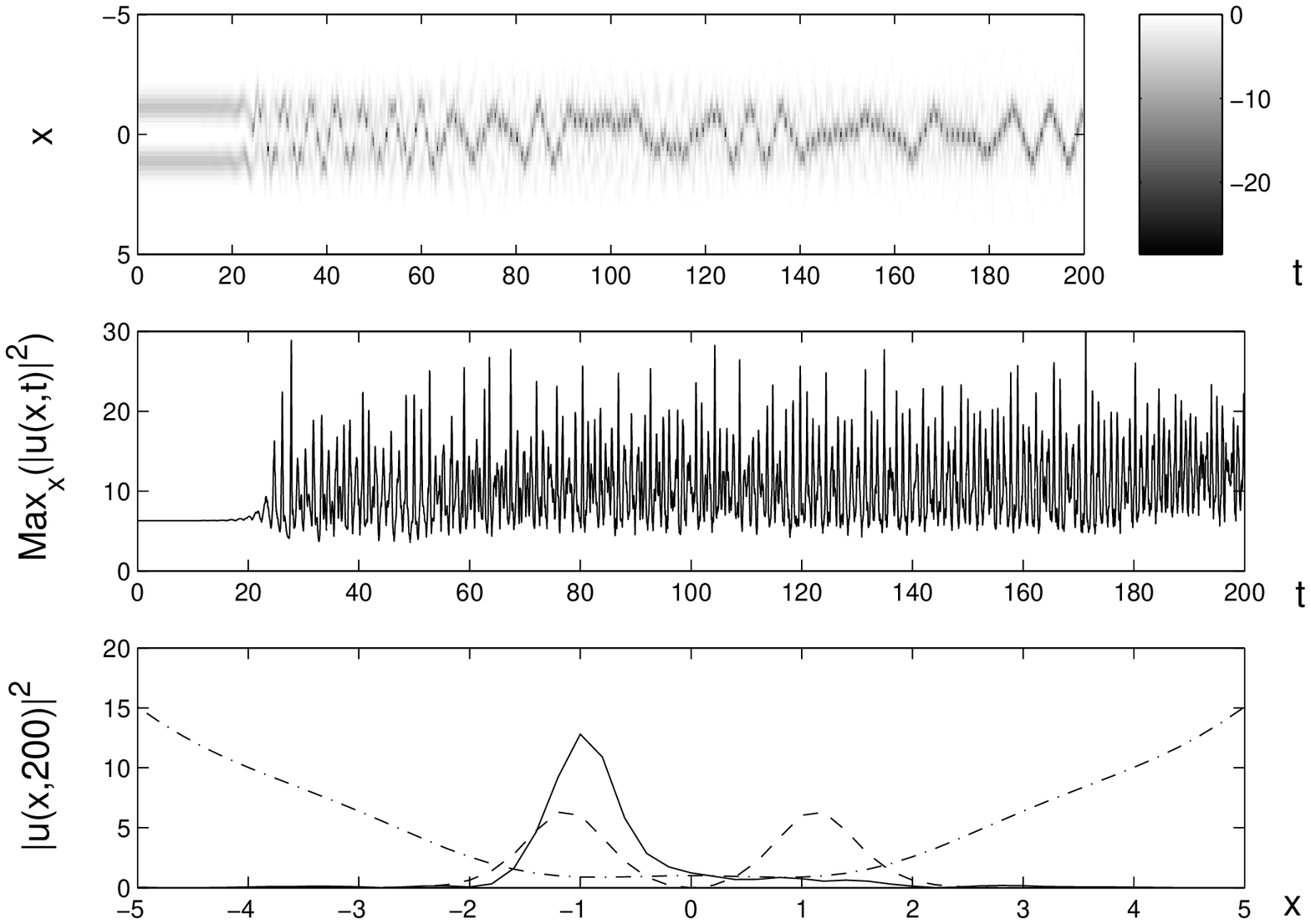}}
\epsfxsize=9cm 
\center{\epsffile{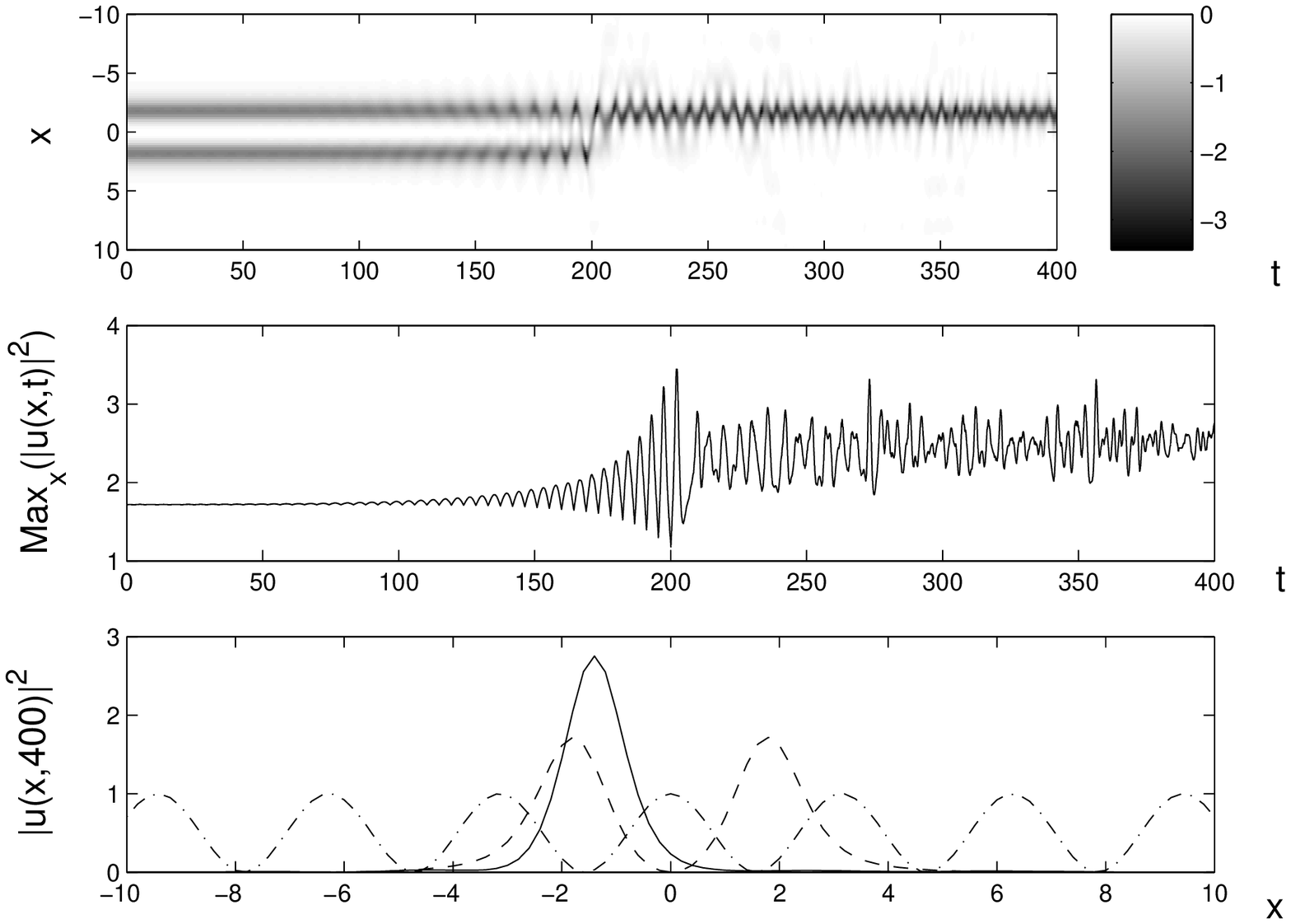}}
\epsfxsize=9cm 
\center{\epsffile{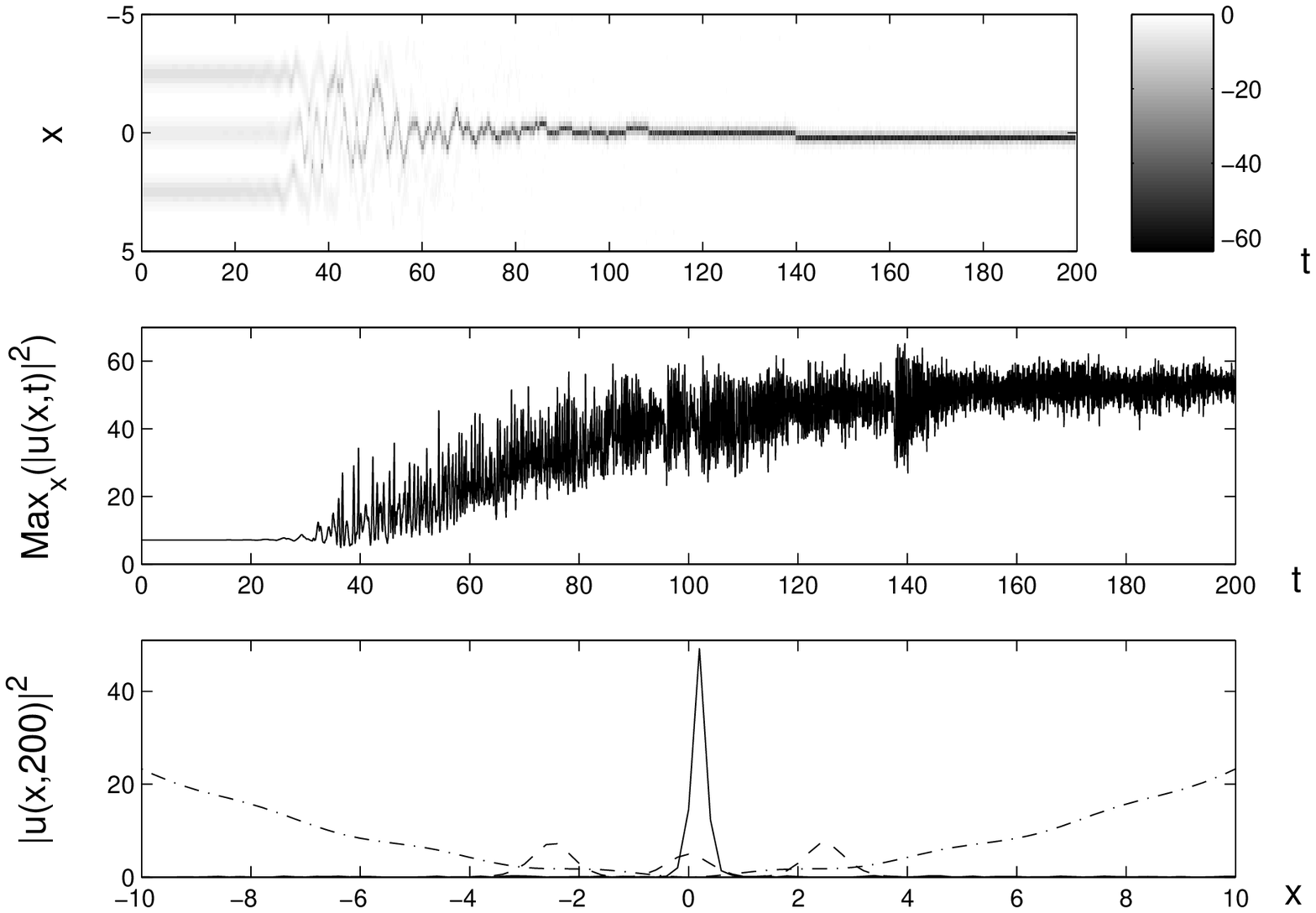}}
\caption{
Typical examples of the directly simulated evolution of unstable TLM-type
solutions. In each set, the top panel displays evolution of the negative
local density, $-|u(x,t)|^{2}$ (using a gray-scale space-time plot),
the middle panel shows the time evolution of the maximum of $|u(x,t)|^{2}$,
and the spatial distribution of the density at the end of the simulation is
presented in the bottom panel. In the latter panel, $|u(x,t)|^{2}$ in final
and initial states, and the potential are shown, respectively, by solid,
dashed, and dashed-dotted lines. The first example is displayed for 
$\epsilon =0.6$ and $V_{0}=1$; recall that the instability growth rate is 
$\approx 0.48$ in this case, i.e., it is a case of strong instability. A
random-noise perturbation of an amplitude $10^{-5}$ was added to the initial
configuration in order to excite the instability. The second example
corresponds to $\epsilon =0$ and $V_{0}=1$ for $\mu =-0.25$. As the
instability growth rate has a very small value in this case, $0.032$, a
random-perturbation seed with a larger amplitude, $\sim 10^{-3}$, was used.
The last example pertains to an unstable bound state of two TLM solitons in
the case of $\epsilon =0.23$ and $V_{0}=1.0$. In this case, a uniformly
distributed random perturbation of amplitude $10^{-5}$ was used to excite
the instability.}
\label{tlmfig5} 
\end{figure}

In the cases for which the TLM solitons or their bound states were found to
be unstable (due to the Hamiltonian Hopf bifurcations), we simulated
nonlinear development of the instability in the framework of the full GP
equation (\ref{tlmeq1}). To this end, a finite-difference scheme with 
$dx=0.2 $ [the same stepsize as in dealing with the stationary equation (\ref
{v})] was used, and time integration was performed by means of the
fourth-order Runge-Kutta scheme with a time step $dt=0.001$. Three examples
of the evolution of unstable configurations are shown in Fig. \ref{tlmfig5}.
The two top ones correspond, respectively, to the strong and weak
oscillatory instability of the TLM soliton. In both of these cases, the TLM
state evolves into a single fundamental soliton. A difference between the
two cases is that, in the former one, with the relatively weak OL, the newly
formed soliton keeps moving between two minima of the potential, while in
the latter case , which corresponds to a stronger OL, the fundamental
soliton becomes pinned. Finally, the bottom subplot of Fig. \ref{tlmfig5}
demonstrates that the oscillatory instability of a bound state of two TLM
solitons transforms it into a single fundamental soliton with a large
amplitude.

\section{Analytical results}

\subsection{The minimum size of the TLM state}

The existence of the TLM state in the GP equation with OL can be explained,
by means of perturbation theory, as a bound state of two fundamental
solitons with phase difference $\pi $ between them. Here, we consider
the case of $\epsilon =0$ (no parabolic trap), but the analysis can be
easily generalized to include the trap. If the present model is considered
as a perturbed NLS equation, we can use an ansatz in the form of Eq. (\ref
{ansatz}) for two solitons. The system's Hamiltonian reads 
\[
H=\int_{-\infty }^{\infty }\left[ |u_{x}|^{2}-\frac{1}{2}|u|^{4}+V_{0}\cos
^{2}\left( \frac{2\pi x}{\lambda }\right) |u|^{2}\right] dx. 
\]
Recall that we have set $\epsilon =0$; here, for generality's sake, we do
not fix $\lambda =2\pi $. Then, known methods (see a review in Ref. \cite
{Progress}) yield an effective net potential of the interaction between two
solitons and interaction of both solitons with the OL. This potential is
realized as a part of the Hamiltonian that depends on the coordinates $\xi
_{1}$ and $\xi _{2}$ of the solitons and phase difference $\Delta \phi $
between them (cf. a similar potential for solitons in the discrete-NLS model
found in Ref. \cite{kkm}) and is of the form: 
\begin{eqnarray}
V_{{\rm eff}}\left( \xi _{1},\xi _{2};\Delta \phi \right) &=&-8\eta ^{3}\cos
\left( \Delta \phi \right) \exp \left( -\frac{\eta }{\sqrt{2}}\left| \xi
_{1}-\xi _{2}\right| \right)  \nonumber \\
&&+\frac{2\sqrt{2}\pi ^{2}V_{0}}{\lambda \sinh \left( \frac{2\sqrt{2}\pi ^{2}
} {\lambda \eta }\right) }\left[ \cos \left( \frac{4\sqrt{2}\pi }{\lambda }
\xi _{1}\right) +\cos \left( \frac{4\sqrt{2}\pi }{\lambda }\xi _{2}\right)
\right] .  \label{Potential}
\end{eqnarray}
The second term in this expression is the Peierls-Nabarro potential (see,
e.g., Ref. \cite{pgk} and references therein) induced by the OL. Equilibrium
positions of the two-soliton system are determined by equations 
\begin{equation}
\frac{\partial V_{{\rm eff}}}{\partial \xi _{1}}=\frac{\partial 
V_{{\rm eff}}} {\partial \xi _{2}}=\frac{\partial V_{{\rm eff}}}
{\partial \left( \Delta
\phi \right) }=0.  \label{equil}
\end{equation}
Straightforward consideration of Eqs. (\ref{equil}) and (\ref{Potential})
shows that a bound state of two solitons with $\Delta \phi =\pi $, which
corresponds to a TLM-like pattern, may exist if the separation $L\equiv
\left| \xi _{1}-\xi _{2}\right| $ between them exceeds the minimum value, 
\begin{equation}
L_{\min }=\frac{\sqrt{2}}{\eta }\ln \left[ \frac{\lambda ^{2}\eta ^{4}} 
{\sqrt{2}\pi ^{3}V_{0}}\sinh \left( \frac{2\sqrt{2}\pi ^{2}}{\lambda\eta}
\right) \right] .  \label{min}
\end{equation}

\begin{figure}[h]
\epsfxsize=9cm 
\center{\epsffile{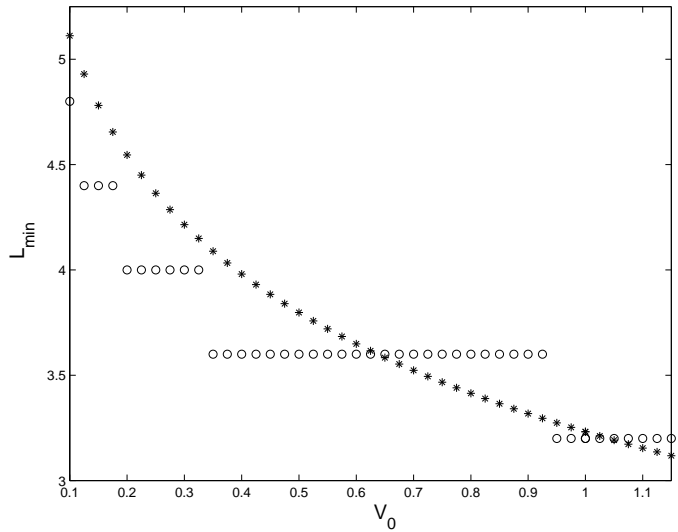}}
\caption{
The analytical prediction (\ref{min}) for the minimum separation between two 
$\pi $-out-of-phase fundamental solitons in the bound state (stars), and the
separation between two density maxima in the TLM found from numerical data
(circles), versus $V_{0}$. Notice that the numerical results should be
considered with an error bar of $\pm 0.2$, due to the discretization used.} 
\label{tlmfig6} 
\end{figure}

This prediction for the minimum separation between fundamental-soliton
components of the TLM state was tested numerically. As a typical example, in
Fig. \ref{tlmfig6} we show the variation of $L_{\min }$ with $V_{0}$ for 
$\lambda =2\pi $ and $\eta =\sqrt{3}$. The distance between the centers of
the solitons was extracted from numerical data as the distance between two
local maxima of $\left| u(x)\right| $, an inherent discretization error of 
$\pm 0.2$ being imposed by the finite-difference scheme. As is seen, the
agreement between the theoretical prediction and the numerically
approximated values of $L_{\min }$ is quite good.

A more detailed analytical consideration of dynamical properties of a
perturbed bound state (following the lines of Ref. \cite{kkm}) shows that,
within the framework of the present approximation, the bound state is
stable. Moreover, the existence of stable bound states of three fundamental
solitons, which correspond to the bound state of two TLMs considered above,
can also be demonstrated by means of this approach.

\subsection{Oscillatory instabilities of the single-TLM and double-TLM 
states}

It is well known (see e.g., Ref. \cite{pgk} and references therein) that the
linearized equations for the perturbations defined in Eq. (\ref{ab}) can be
rewritten, in terms of new variables $U=a+b^{\ast }$ and $W=a-b^{\ast }$, as 
\begin{eqnarray}
\omega U &=&L_{-}W\equiv -W_{xx}-v^{2}W+(V(x)-\mu )W,  \label{keq6} \\
\omega W &=&L_{+}U\equiv -U_{xx}-3v^{2}U+(V(x)-\mu )U_{n}.  \label{keq7}
\end{eqnarray}
The {\it Krein signature} of the eigenvalue can be written as \cite{JohAub} 
\[
K={\rm sign}\left( -\int_{-\infty }^{\infty }(UW^{\star }+WU^{\star
})dx\right) . 
\]
Analytical consideration is possible for $\omega $'s that are real; then $U$
and $W$ are also real, hence $K$ amounts to the expression 
\begin{equation}
K={\rm sign}\left( -\int_{-\infty }^{\infty }UWdx\right) .  \label{Krein}
\end{equation}

As was done above, we approximate a TLM solution as a superposition, 
$v(x)=P(x)+N(x)$, of two stationary fundamental-soliton waveforms, positive 
$P(x)$ and negative $N(x)$, the underlying assumption being a weak overlap
between $P(x)$ and $N(x)$. Since $P$ and $N$ are, essentially, individual
solitons, in the lowest-order approximation they obey equations $\left[
V(x)-\mu \right] P=P_{xx}+P^{3}$ and $\left[ V(x)-\mu \right] 
N=N_{xx}+N^{3}$. Summing them up, and taking into account 
that the product $\left|
PN\right| $ is everywhere much smaller than $P^{2}+N^{2}$ (due to the
assumed weak overlap between $P$ and $N$), we conclude that, up to the same
lowest-order accuracy, the function $P(x)+N(x)$ satisfies the equation 
$L_{-}(P+N)=0$ [recall that the operator $L_{-}$ is defined in Eq. 
(\ref{keq6})]. Thus, in the lowest approximation, $P(x)+N(x)$ 
is a zero mode of the
operator $L_{-}$. In a similar way, the following approximate result can be
obtained for the function $P(x)-N(x)$: 
\begin{equation}
L_{-}(P-N)=-PN(P-N).  \label{LPN}
\end{equation}

The linearization around each of the TLM-constituent pulses (if considered
in isolation) carries a pair of zero eigenvalues due to the phase
invariance. The corresponding eigenfunction, $W$, has the shape of the pulse
itself ($P$ and $N$, respectively). There is also a generalized
eigenfunction \cite{JohAub} for each pair. We expect to have two pairs of
TLM eigenvalues originating from these pairs of the individual-soliton's
ones, the corresponding eigenfunctions being (in the lowest approximation) 
$W=P+N$ and $W=P-N$. The former one is related to the phase invariance of the
system (in other words, to the conservation of the total number of atoms),
which means that the corresponding eigenvalue pair remains exactly equal to
zero, while the other one may become different than zero. Then, it follows
from Eq. (\ref{LPN}) and Eq. (\ref{keq6}) that, in the lowest approximation, 
$U=-PN(P-N)$. In turn, Eq. (\ref{Krein}) yields

\begin{equation}
K={\rm sign}\left[ \int_{-\infty }^{\infty }PN(P-N)^{2}dx\right] .
\label{keq16}
\end{equation}
Since, by definition, the signs of $P(x)$ and $N(x)$ are opposite, Eq. (\ref
{keq16}) implies that the Krein signature of this phase eigenmode is
negative. A known consequence of this \cite{VdM} is that an oscillatory
instability will be generated by collision of this nonzero eigenvalue with
other discrete ones (and/or those belonging to the continuous spectrum) that
carry a positive Krein signature.

The above consideration indicates that, for every pair of the interacting 
$\pi $-out-of-phase pulses, there exists an eigenvalue with  negative
Krein signature and, consequently, a potentially ensuing Hamiltonian Hopf
bifurcation. This conclusion fully agrees with our numerical results
presented above, that have identified a single oscillatory instability for
the single TLM soliton, and {\em two} such instabilities in the case of
bound states of two TLM solitons.

\subsection{TLM solitons in the absence of the OL}

Finally, it is possible to present analytical arguments justifying the
existence of TLM solitons (both single and multiple ones) even when the
harmonic trap is overwhelmingly stronger than the OL potential (see the
numerical results shown in the last panels in Figs. \ref{tlmfig2} and \ref
{tlmfig4}), or when the OL potential is simply absent. Within the framework
of the same perturbation theory which was employed above to derive Eq. (\ref
{Potential}), each of the pulses constituting the TLM state is then subject
to the action of two forces. One of them is exerted by the magnetic trap,
while the other comes from the interaction between the pulses. Accordingly,
the effective potential is 
\begin{eqnarray}
V_{{\rm eff}}(\xi _{1},\xi _{2};\Delta \phi ) &=&-8\eta ^{3}\cos \left(
\Delta \phi \right) \exp \left( -\frac{\eta }{\sqrt{2}}\left| \xi _{1}-\xi
_{2}\right| \right)  \nonumber \\
&&+4\sqrt{3}\epsilon \left( \xi _{1}^{2}+\xi _{2}^{2}\right) .
\label{tlmeq7}
\end{eqnarray}

\begin{figure}[h]
\epsfxsize=9cm 
\center{\epsffile{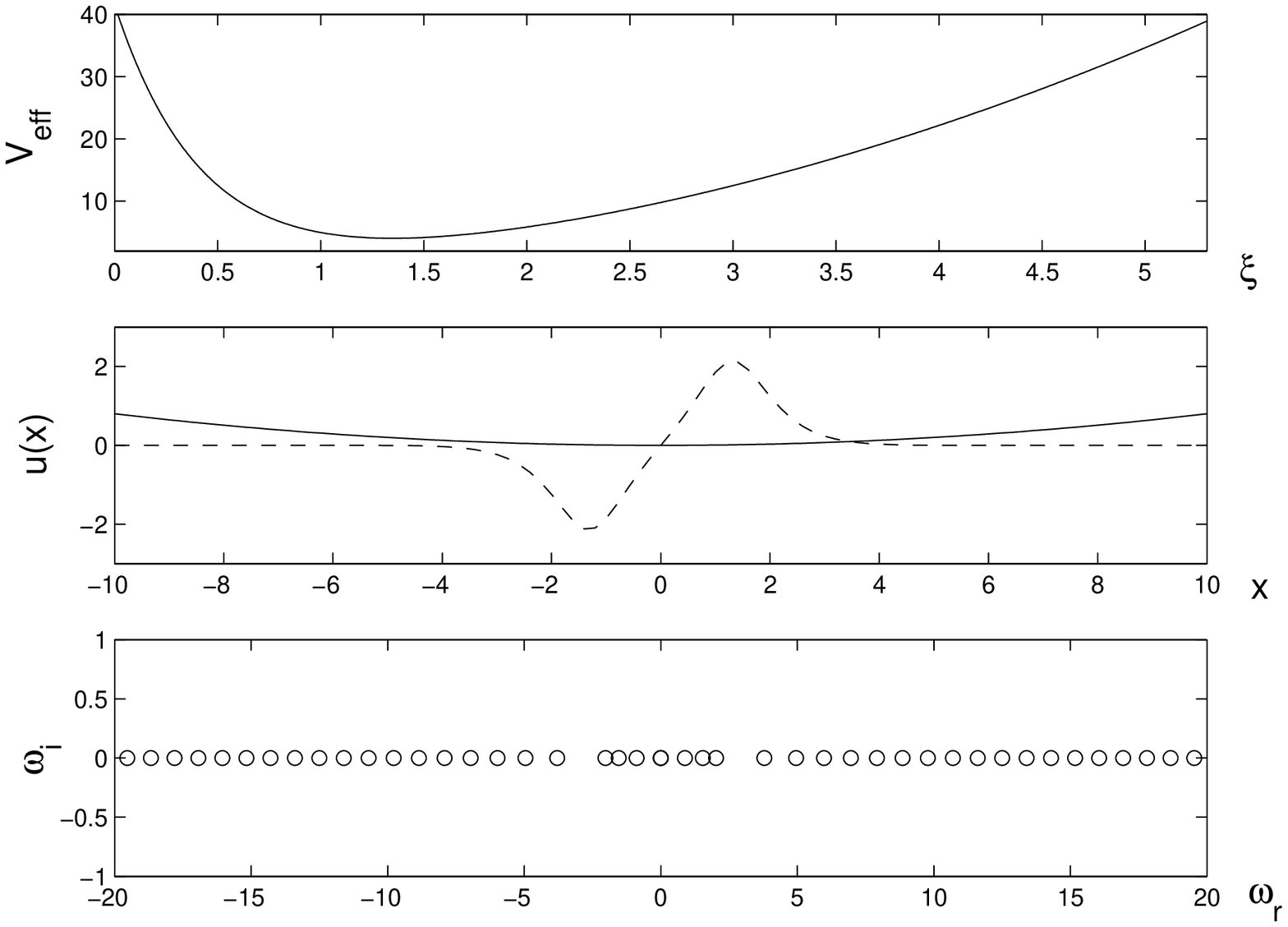}}
\caption{
The top panel shows the effective potential $V_{{\rm eff}}$, as a function
of the coordinate $\xi $ of the right pulse, which is induced by the
repulsion between the pulses and the restoring force from the magnetic trap,
following the analytical expression (\ref{tlmeq7}). The potential minimum
(equilibrium position) is located at $\xi \approx 1.35$. The middle and
bottom panels show the TLM configuration and its stability, as found in the
numerical form. The equilibrium position corresponding to the numerical
solution is inferred from the middle plot to be at $\xi \approx 1.4$. The
parameters are $V_{0}=0$, $\epsilon =0.2$, and $\mu =-1.5$.}
\label{tlmfig7} 
\end{figure}

Predictions following from this approximation were tested against direct
numerical results obtained in the absence of the OL, i.e., with $V_{0}=0$ in
Eq. (\ref{tlmeq1}). In Fig. \ref{tlmfig7}, an example is displayed for 
$\epsilon =0.2$ and $\mu =-1.5$. The top panel shows the effective potential 
(\ref{tlmeq7}), which, in this case, has a minimum at $\xi _{1,2}=\pm 1.35$.
The middle and bottom panels present a numerical solution, with maxima of 
$\left| u(x)\right| $ at $\xi _{1,2}\approx \pm 1.4$. Thus, the perturbative
approximation not only explains the existence of the TLM state in attractive
BECs even in the absence of OL, but also accurately predicts the location of
centers of the pulses whose concatenation generates the TLM structure. This,
in turn, supports the numerical findings\ presented in section II, which
suggested that TLM solitons in the attractive BEC would indeed persist even
for very tight magnetic traps. Thus, we conclude that the TLM solitons are a
very robust feature of the system.

\section{Conclusions}

In this work we have demonstrated that counterparts of twisted localized
modes (TLMs), which were earlier found in the discrete one-dimensional NLS
equation, can exist as robust objects in attractive, effectively
one-dimensional Bose-Einstein condensates (BECs), provided that the
condensate is confined by the optical lattice (OL) and/or parabolic magnetic
trap. The TLMs in BECs may be considered as bound states of two fundamental
solitons with a phase difference $\pi $ between them, or, alternatively, as
a dark soliton embedded in a broader bright one. The family of the TLM
solitons and their stability were investigated in detail, by means of
numerical continuation in the strengths of the OL or magnetic-trap
parameters, as well as in the chemical potential. Stable bound states of two
TLMs, alias a bound state of three fundamental solitons with a
$\pi$ phase shift 
between adjacent ones, were also found. In the cases when the TLMs
are unstable, the development of the instability was monitored by means of
direct simulations, with a conclusion that the TLM rearranges itself into a
fundamental soliton.

We have also developed an analytical approach, which treats each fundamental
soliton as a quasi-particle subject to the action of effective forces
exerted by the OL, magnetic trap, and interaction with another soliton. The
analysis predicts the existence of the stable bound state of two fundamental
solitons with a phase shift of $\pi $ between them. The smallest possible
distance between the bound solitons, predicted by this approach, is in good
agreement with the numerically found size of the corresponding TLM. The
analysis can be readily extended to predict that multi-pulse bound states
with the phase shift $\pi $ between adjacent solitons also exist and may be
stable.

The origin of the instability of the TLMs, in the case when they are
unstable, was also examined by means of an analytical approximation, and was
found to stem from Hamiltonian Hopf bifurcations, which occur when
eigenvalues with negative Krein signature collide with
positive-signature ones. This result was based on the existence of a
negative-Krein-sign eigenvalue set, that was constructed by means of the
perturbative analysis.

It will be very interesting to generalize these considerations to higher
dimensions -- in particular to two-dimensional settings. In that case, a
relevant issue is the possible existence of a 2D counterpart of the TLM in the
form of a bright vortex (which was found in the 2D discrete NLS equation,
where it was shown to be stable under certain conditions \cite{dv}). A very
recent preliminary result is that bright vortices, stabilized by a 2D
optical lattice, can indeed be found \cite{Salerno}.

\section*{Acknowledgements}

One of the authors (B.A.M.) appreciates valuable discussions with B.
Baizakov and M. Salerno, and hospitality of the Department of Physics at
Universit\'{a} di Salerno (Italy). The work of this author was supported, in
a part by a Window-on-Science grant from the European Office of Aerospace
Research and Development (EOARD) of US\ Air Force. This project was also
supported by a University of Massachusetts Faculty Research Grant and
NSF-DMS-0204585 (P.G.K.), the Special Research Account of the University of
Athens (D.J.F.), the Binational (US-Israel) Science Foundation, under grant
No. 1999459 (B.A.M.) and by the AFOSR Dynamics and Control (I.G.K.). Work at
Los Alamos is supported by the US DoE.



\begin{references}

\bibitem{review}  F. Dalfovo, S. Giorgini, L.P. Pitaevskii, and S.
Stringari, Rev. Mod. Phys. {\bf 71}, 463 (1999). 

\bibitem{dark}  S. Burger {\it et al.}, Phys. Rev. Lett. {\bf 83},
5198(1999); J. Denschlag {\it et al.}, Science {\bf 287}, 97 (2000); B.P.
Anderson {\it et al.}, Phys. Rev. Lett. {\bf 86}, 2926 (2001).

\bibitem{bright}  K.E. Strecker, G.B. Partridge, A.G. Truscott, and R.G.
Hulet, Nature {\bf 417}, 150 (2002); L. Khaykovich, F. Schreck, G. Ferrari,
T. Bourdel, J. Cubizolles, L.D. Carr, Y. Castin, and C. Salomon, Science 
{\bf 296}, 1290 (2002).

\bibitem{vortex}  M.R. Matthews {\it et al.}, Phys. Rev. Lett. {\bf 83},
2498(1999); K.W. Madison, {\it et al.} Phys. Rev. Lett. {\bf 84}, 806
(2000); S. Inouye {\em et al.}, Phys. Rev. Lett.{\bf {87}}, 080402 (2001).


\bibitem{stal}  K. Staliunas, S. Longhi and G. J. de Valc\'{a}rcel, 
\newblock
Phys. Rev. Lett. {\bf 89}, 210406 (2002).

\bibitem{theo}  G. Theocharis, D.J. Frantzeskakis, P.G. Kevrekidis, B.A.
Malomed and Yu.S. Kivshar, ``Ring dark solitons and vortex necklaces in
Bose-Einstein condensates'', e-print cond-mat/0302102; Phys. Rev. Lett., in
press.

\bibitem{inouye}  S. Inouye {\it et al.}, Nature {\bf 392}, 151 (1998); E.A.
Donley {\it et al.}, Nature {\bf 412}, 295 (2001).

\bibitem{louis}  P.J.Y. Louis, E.A. Ostrovskaya, C.M. Savage and Yu.S.
Kivshar, Phys. Rev. A {\bf 67}, 013602 (2003).

\bibitem{DKL}  S. Darmanyan, A. Kobyakov and F. Lederer, \newblock Sov.
Phys. JETP {\bf 86}, 682 (1998); P.G. Kevrekidis, A.R. Bishop and K.{\O }.
Rasmussen, \newblock Phys. Rev. E {\bf 63}, 036603 (2001).

\bibitem{Li}  A.J. Moerdijk, W.C. Stwalley, R.G. Hulet, and B.J. Verhaar,
Phys. Rev. Lett. {\bf 72}, 40 (1994).

\bibitem{Ru}  J.P. Burke, J.L. Bohn, B.D. Esry, and C.H. Greene, Phys. Rev.
Lett. 80, 2097 (1998).

\bibitem{vvkbbms}  B.B. Baizakov, V.V. Konotop and M. Salerno, \newblock J.
Phys. B 35, 5105 (2002); K.M. Hilligsoe, M.K. Oberthaler, and K.P. Marzlin,
Phys. Rev. A {\bf 66}, 063605 (2002); 
E.A. Ostrovskaya and Y.S. Kivshar, e-print cond-mat/0303190 (2003).

\bibitem{rupr}  P.A.\ Ruprecht, M.J.\ Holland, K.\ Burnett and M.\ Edwards,
\newblock Phys.\ Rev.\ A {\bf 51}, 4704 (1995).

\bibitem{GPE1d}  V.M. P\'{e}rez-Garc\'{i}a, H. Michinel and H. Herrero,
Phys. Rev. A {\bf 57}, 3837 (1998); L. Salasnich, A. Parola and L.
Reatto,Phys. Rev. A {\bf 65}, 043614 (2002); Y.B. Band, I. Towers, and B.A.
Malomed, Phys. Rev. A {\bf 67}, 023602 (2003).

\bibitem{catal}  F.S.\ Cataliotti, S.\ Burger, C.\ Fort, P.\ Maddaloni, F.\
Minardi, A.\ Trombettoni, A.\ Smerzi and M.\ Inguscio, \newblock Science 
{\bf 293}, 843 (2001).

\bibitem{greiner}  M.\ Greiner, I.\ Block, O.\ Mandel, T.W.\ H{\"{a}}nsch
and T.\ Esslinger, \newblock Appl.\ Phys.\ B {\bf 73}, 769 (2001).

\bibitem{tromb}  A.\ Trombettoni and A.\ Smerzi, Phys. Rev. Lett. {\bf 86},
2353, (2001).

\bibitem{konot}  F.Kh.\ Abdullaev, B.B.\ Baizakov, S.A.\ Darmanyan, V.V.\
Konotop, and M.\ Salerno, Phys.\ Rev.\ {\bf A64}, 043606 (2001).

\bibitem{tromb2}  A.\ Smerzi, A.\ Trombettoni, P.G.\ Kevrekidis, and A.R.\
Bishop Phys.\ Rev.\ Lett.\ {\bf 89}, 170402 (2002).

\bibitem{catal2}  F. S. Cataliotti, L. Fallani, F. Ferlaino, C. Fort, P.
Maddaloni, M. Inguscio, A. Smerzi, A. Trombettoni, P. G. Kevrekidis, A. R.
Bishop, ``A Novel mechanism for superfluidity breakdown in weakly coupled
Bose-Einstein condensates'', e-print cond-mat/0207139.


\bibitem{findif} For finite difference schemes for partial differential
equations the interested reader can be referred to e.g.,
G.D. Smith,  {\it Numerical Solution of Partial Differential Equations:
Finite Difference Methods}, Oxford
University Press (Oxford, 1986); G.E. Forsythe and W.R. Wasow,
{\it Finite-Difference Methods for Partial Differential Equations},
John Wiley \& Sons, Inc. (New York, 1960).
For Newton type methods, as well as finite difference schemes, one
can consult: W.H. Press {\it et al.}, {\it Numerical Recipes in C: 
The Art of Scientific Computing}, Cambridge University Press
(Cambridge, 1988).

\bibitem{kkm}  T. Kapitula, P.G. Kevrekidis and B.A. Malomed, \newblock
Phys. Rev. E {\bf 63}, 036604 (2001).

\bibitem{jared}  J.C. Bronski {\it et al.} Phys. Rev. Lett. {\bf 86}, 1402
(2001); J.C. Bronski {\it et al.} Phys. Rev. E {\bf 64}, 056615 (2003).

\bibitem{matlab} {\it Matlab, The Language of Scientific Computing},
the Mathworks Inc., Natick, MA (2000).

\bibitem{pgk}  P.G. Kevrekidis, K.\O . Rasmussen and A.R. Bishop, Int. J.
Mod. Phys. B {\bf 15}, 2833 (2001).

\bibitem{JohAub}  M. Johansson and S. Aubry, \newblock Phys. Rev. 
E {\bf 61}, 5864 (2000).

\bibitem{VdM}  J.-C. van der Meer, \newblock Nonlinearity {\bf 3}, 1041
(1990); see also e.g., I.V. Barashenkov, D.E. Pelinovsky and E.V.
Zemlyanaya, \newblock 
Phys. Rev. Lett. {\bf 80}, 5117 (1998); \newblock A. De Rossi, C. Conti and
S. Trillo, \newblock Phys. Rev. Lett. {\bf 81}, 85 (1998); M. Johansson and
Yu.S. Kivshar, \newblock Phys. Rev. Lett. {\bf 82}, 85 (1999).

\bibitem{Progress}  B.A. Malomed, Progr. Optics {\bf 43}, 71 (2002).





\bibitem{dv}  B.A. Malomed and P.G. Kevrekidis, {Phys. Rev. E} {\bf 64},
026601 (2001).

\bibitem{Salerno}  B. Baizakov, B.A. Malomed, and M. Salerno, in preparation.

\end{references}
\end{document}